\documentclass{elsarticle}
\usepackage{algorithmic}
\usepackage{algorithm}
\usepackage{multirow}
\usepackage{stfloats}
\usepackage{caption}
\usepackage{subcaption}
\usepackage{booktabs}
\usepackage{amsmath}

\AtBeginDocument{%
  }

\begin{document}
\begin{frontmatter}
\title{Scalable FPGA Framework for Real-Time Denoising in High-Throughput Imaging: A DRAM-Optimized Pipeline using High-Level Synthesis}

\author[1]{Weichien Liao}
\ead{liaowei2@msu.edu}
\affiliation[1]{%
  organization={Department of Computer Science and Engineering, Michigan State University},
  city={East Lansing},
  country={USA}}




\begin{abstract}
    High-throughput experimental workflows, such as Parallel Rapid Imaging with Spectroscopic Mapping (PRISM), generate imaging data at rates exceeding conventional real-time processing capabilities. We present a hardware-accelerated preprocessing pipeline for data denoising based on Field-Programmable Gate Arrays (FPGAs), using High-Level Synthesis (HLS) with DRAM-backed buffering. Our design performs frame subtraction and averaging directly on streamed image data, minimizing DRAM access latency through burst-mode AXI4 interfaces. The resulting kernel operates below the inter-frame interval, enabling real-time performance and significantly reducing dataset size for more efficient downstream analysis on CPU or GPU platforms. This modular, scalable FPGA implementation offers a practical solution for latency-sensitive imaging workflows, particularly in spectroscopic and microscopy applications.
\end{abstract}


\end{frontmatter}

\section{Introduction and Motivation}

The widespread adoption of high-throughput experimental techniques in materials science \cite{Luo2023, Kezilebieke2020, Zhang2017} , biology \cite{Frei2025, Chen2021, Rosenfeld2011, Wang2019, Wu2017} and medical imaging \cite{Mitani2018} has ushered in a new era of rapid data generation. Advanced imaging modalities---particularly high-resolution cameras used in microscopy and spectroscopy---produce data streams at rates that often overwhelm conventional processing pipelines. This surge in data volume and structural complexity creates a critical bottleneck, impeding real-time analysis, experimental feedback, and timely extraction of scientifically meaningful insights. To mitigate these challenges, automated and latency-aware processing methods are essential to extract relevant information in real time, enabling dynamic signal interpretation and informed control during ongoing experiments.

While general-purpose CPUs and GPUs provide substantial computational power, they often fail to meet the stringent latency and throughput demands of real-time processing for high-throughput imaging workflows. CPUs, inherently optimized for sequential tasks, offer limited parallelism and struggle to keep pace with continuous data streams. GPUs, though massively parallel, introduce performance trade-offs—particularly due to data transfer overheads between acquisition hardware and GPU memory. These transfers, coupled with inefficient handling of irregular data access patterns, create latency bottlenecks that are especially detrimental in applications requiring immediate processing and low-latency feedback.

Field-Programmable Gate Arrays (FPGAs) offer a compelling alternative for real-time data processing, combining deep hardware-level parallelism with low-latency execution and reconfigurable architecture. Their integration into image-capturing hardware is facilitated by relatively low cost and compact form factor, making them well-suited for edge deployment. Crucially, FPGAs can implement custom accelerators tailored to specific algorithms and dataflow patterns---an architectural advantage for managing the intensive computational demands of high-throughput experiments. Unlike CPU or GPU solutions that rely on general-purpose software pipelines, FPGA implementations execute logic directly in hardware, enabling true real-time processing of streaming datasets.

Despite their real-time processing potential, traditional FPGA programming with hardware description languages (HDLs) such as VHDL and Verilog poses significant barriers to adoption. These languages demand expertise in digital circuit design, including clock timing, pipelining, and concurrency management—making development and debugging both intricate and time-intensive. High-Level Synthesis (HLS) tools, such as AMD Vitis HLS, help overcome these limitations by enabling designers to describe hardware behavior using high-level languages like C or C++. The compiler automatically converts these abstractions into optimized HDL implementations, dramatically accelerating development while preserving performance. Crucially, HLS frameworks offer configurable directives that empower non-specialists to produce efficient, application-specific designs, broadening access to FPGA-based acceleration across scientific domains.

In this work, we investigate the use of FPGAs to enable real-time processing of image data generated in high-throughput experimental workflows, with a particular emphasis on high-speed camera acquisition. We introduce a generalizable framework for FPGA-based image processing pipelines built using Vitis High-Level Synthesis (HLS). The architecture addresses memory access bottlenecks by efficiently utilizing on-board DRAM—essential for workflows involving delayed frame access or large image resolutions. We demonstrate the performance of this framework in the context of Parallel Rapid Imaging with Spectroscopic Mapping (PRISM) \cite{Wu2025}, where the FPGA is embedded directly within the acquisition hardware. This integration enables real-time analysis and feedback control, facilitating the timely exploration of subtle or transient physical phenomena.

The remainder of the paper is organized as follows. Section~\ref{sec:background} outlines the PRISM workflow and its underlying motivation. Section~\ref{sec:related} reviews related research in real-time image processing and FPGA acceleration. Section~\ref{sec:methods} details our algorithmic approach and FPGA implementation for high-performance subtraction and averaging. Section~\ref{sec:experiment} describes the experimental setup used to evaluate the framework under realistic PRISM imaging conditions. Section~\ref{sec:hls-performance} reports simulation results from Vitis HLS demonstrating latency performance, followed by Section~\ref{sec:fpga-perform}, which compares execution characteristics across CPU, GPU, and FPGA platforms. Section~\ref{sec:conclusion} concludes with a summary of key findings and discusses implications for future work in scalable real-time imaging pipelines.

\section{PRISM Workflow: Background and Imaging Challenges} \label{sec:background}
\begin{figure}[H]
    \centering
    \includegraphics[width=0.5\linewidth]{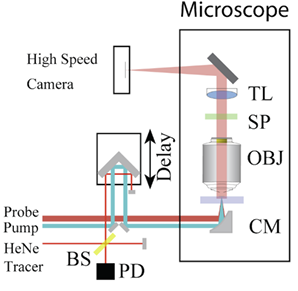}
    \caption{Schematic overview of the PRISM experimental setup used for ultrafast spectroscopic imaging. BS: Beam Splitter; PD: Photodiode; CM: Curved Mirror; OBJ: Objective; SP: Short pass Filter; TL: Tube Lens. The diagram illustrates beam paths, optical elements, and synchronization components, including the collinear pump–probe configuration and detection system interfaced with a high-speed camera.}
    \label{fig:prism-setup}
\end{figure}

Parallel Rapid Imaging with Spectroscopic Mapping (PRISM) \cite{Wu2025} is an ultrafast coherent imaging technique engineered to capture vibrational and electronic dynamics with femtosecond temporal resolution and sub-diffraction spatial precision. Utilizing wide-field pump–probe microscopy, PRISM generates high-throughput spectroscopic image streams across nearly $100,000$ pixels. Its acquisition pipeline employs a collinear pump–probe configuration, combining a second harmonic optical parametric amplifier (2H NOPA) as the pump and a third harmonic (3H NOPA) as the probe. These beams are spectrally tuned for optimal excitation and signal isolation: the pump initiates molecular dynamics while the probe captures their evolution. The system synchronizes a high-speed camera with a voice-coil stage to acquire up to $1.6$ million wideband spectra per second (spanning $5$–$600$ cm$^{-1}$), producing dense 3D datasets in spatial $(x, y)$ and temporal ($\Delta t$) domains. Figure~\ref{fig:prism-setup} provides a schematic overview of the PRISM experimental setup, including beam paths, optical elements, and synchronization components.

To extract meaningful signals from raw PRISM datasets, the \textbf{data analysis stage} applies several advanced post-processing techniques. Time-resolved image sequences undergo multi-exponential curve fitting to isolate population decay dynamics from oscillatory behavior. Fast Fourier Transforms (FFT) are then applied to the residuals, generating hyperspectral maps that capture underlying vibrational modes. To further suppress noise and enhance contrast, singular value decomposition (SVD) is employed for dimensionality reduction and signal refinement.

To enhance the signal-to-noise ratio (SNR), PRISM performs averaged scans by capturing multiple pump–probe delay sequences with a high-speed camera operating at $20{,}000$ frames per second (FPS), while modulating the delay dynamically via a $10$ Hz voice coil stage. Each complete scan spans $50$ ms, and multiple consecutive scans---typically $10$ for a $1$-second acquisition---are recorded and processed. This acquisition scheme yields a sequence of alternating scans, with and without excitation, corresponding to different molecular states. Following acquisition, scans undergo frame-by-frame subtraction to isolate excitation-induced signal components, followed by averaging to further suppress random noise and improve both spectral and temporal resolution. This procedure yields more precise measurements of decay constants, vibrational frequencies, and other signal features, particularly those too weak or transient to be reliably captured in a single scan. In PRISM’s computational workflow, this subtraction-and-averaging step forms a dedicated \textbf{preprocessing stage} preceding the downstream analysis techniques described above. The following section reviews related work in FPGA-based data processing pipelines and streaming architectures for spectroscopic and microscopic imaging.




\section{Related Work in FPGA Imaging Pipelines}\label{sec:related}
A 2011 study \cite{Montgomery2011} introduced a real-time 4D microscopy system using a high-speed camera connected to an FPGA, implementing two fringe detection algorithms directly in hardware. These algorithms operated exclusively on the current frame, eliminating the need for on-chip or off-chip memory access and resulting in no emphasis on BRAM or DRAM utilization. Another effort \cite{Zhu2022} proposed an FPGA accelerator for filtering, detection, and tracking of motion in spike camera imagery, referring to asynchronous event-driven pixel data captured by neuromorphic sensors. However, similar to the earlier work, this application did not require retention or access to prior frames, limiting its relevance for workflows with deep temporal dependencies.


Several additional studies have examined FPGA integration in workflows involving high-speed camera data. One work \cite{Binotto2013} proposes an on-site X-ray imaging pipeline using a heterogeneous system composed of CPU, GPU, and FPGA components. Another approach \cite{Vellas2017} employs FPGA acceleration for similarity-based matching of spectral signatures in hyperspectral data. The design presented in \cite{Sarkar2021} leverages an embedded FPGA within the camera itself for format conversion and on-board control in live video streaming. Further contributions explore FPGA implementations for real-time data compression \cite{Regorsek2024} and convolutional neural network (CNN)-based mode tracking \cite{Wei2024}, highlighting the growing utility of reconfigurable hardware for low-latency image processing.


Additional studies \cite{Gyaneshwar2022, Gonzalez2016, Nascimento2020, Safaei2016} have focused on using FPGAs to accelerate data processing in scientific workflows, including hyperspectral image classification, spectral target detection, and 4D surface analysis. These examples reinforce the applicability of reconfigurable hardware beyond conventional imaging tasks, particularly in domains requiring low-latency computation and large data throughput. However, many of these implementations operate on small working sets or rely exclusively on on-chip memory, limiting their scalability. This underscores the gap addressed in Section~\ref{sec:methods}, where we introduce a DRAM-integrated FPGA framework capable of processing large-scale imaging data streams in real time.



\section{FPGA Preprocessing Architecture and Algorithms} \label{sec:methods}
As discussed in Section~\ref{sec:related}, prior FPGA-based implementations for real-time image processing often operate on small datasets involving only single frames or short frame sequences. These approaches typically leverage Block Random Access Memory (BRAM)---on-chip memory that runs at FPGA clock speed and offers nanosecond-level access latency---for data storage and retrieval. While BRAM enables low-latency performance, its limited capacity (ranging from kilobytes to a few megabytes) restricts scalability, especially for workflows requiring access to large volumes of image data. On-board DRAM provides higher-capacity storage but incurs significantly greater latency, on the order of tens of nanoseconds, complicating efforts to maintain real-time throughput. \emph{The framework we propose addresses this challenge by achieving real-time performance on large working sets using DRAM-backed buffering and burst-mode AXI4 data transfers.} Although demonstrated in the context of PRISM, the architecture is modular and implemented via AMD Vitis High-Level Synthesis (HLS), making it adaptable to other high-throughput imaging workflows.


As detailed in Section~\ref{sec:background}, the PRISM workflow consists of three primary stages, with this paper focusing on computations in the \textbf{preprocessing stage}. In the experimental setup (see Section~\ref{sec:experiment}), image data acquired by a high-speed camera are continuously streamed into a frame grabber equipped with an embedded FPGA and on-board DRAM. This configuration makes it both practical and essential to integrate preprocessing directly into the \textbf{data acquisition pipeline}. Critically, the preprocessing stage performs early denoising and dimensionality reduction, substantially shrinking the dataset before transmission to downstream analysis modules. As such, it represents a central computational bottleneck in realizing real-time performance within PRISM’s end-to-end pipeline.


\subsection{Subtraction and Averaging Strategy}

\begin{figure}
    \centering
    \includegraphics[width=1\linewidth]{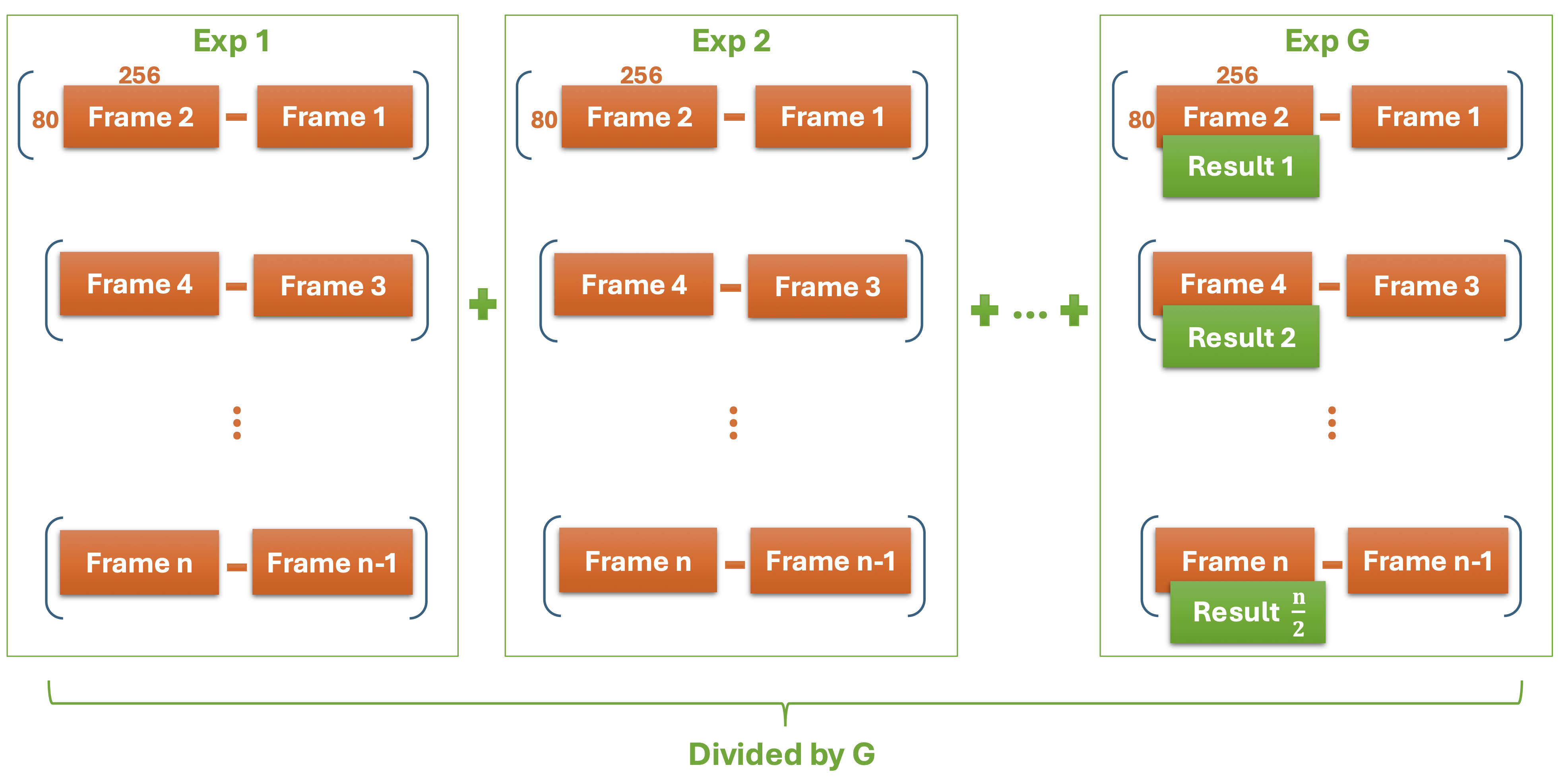}
    \caption{Frame-wise subtraction and averaging schema used during PRISM preprocessing. For each experiment, sequential scans generate alternating excitation and control frames. Subtraction between consecutive frames isolates excitation-induced signals, which are subsequently averaged across all groups to suppress random noise and enhance signal fidelity. Each image result is computed as the groupwise average of difference frames and represents the final denoised output.}
    \label{fig:prism-preprocess}
\end{figure}

Figure~\ref{fig:prism-preprocess} illustrates the core computations performed during the preprocessing stage. In this scenario, $G$ experiments are executed sequentially, yielding $G$ groups of raw image frames. Each group consists of $N$ frames, where $N$ is even, and every frame has fixed dimensions of height ($H$) and width ($W$), forming images of size $H\times W$ pixels. As described earlier, PRISM data exhibit an alternating pattern---frames with and without the signal of interest---enabling pixel-wise subtraction between consecutive frames to isolate meaningful signal contributions. To further suppress random noise and enhance both spectral and temporal precision, the resulting difference frames are averaged across each group. Although these operations are conceptually straightforward, their scale presents a significant implementation challenge: frame rates often exceed several thousand per second, $N$ can reach thousands of frames, and $G$ may span tens of groups. As a result, the core difficulty lies in designing FPGA algorithms that respect memory constraints while efficiently utilizing the AXI4-stream interface to access on-board DRAM and maintain real-time throughput. To ensure the validity of the computational argument, we assume that the noise sequence is a stationary process, the signal of interest consists of i.i.d. random variables, and the noise and signal components are uncorrelated.

\begin{figure}[H]
    \centering
    \includegraphics[scale=0.08]{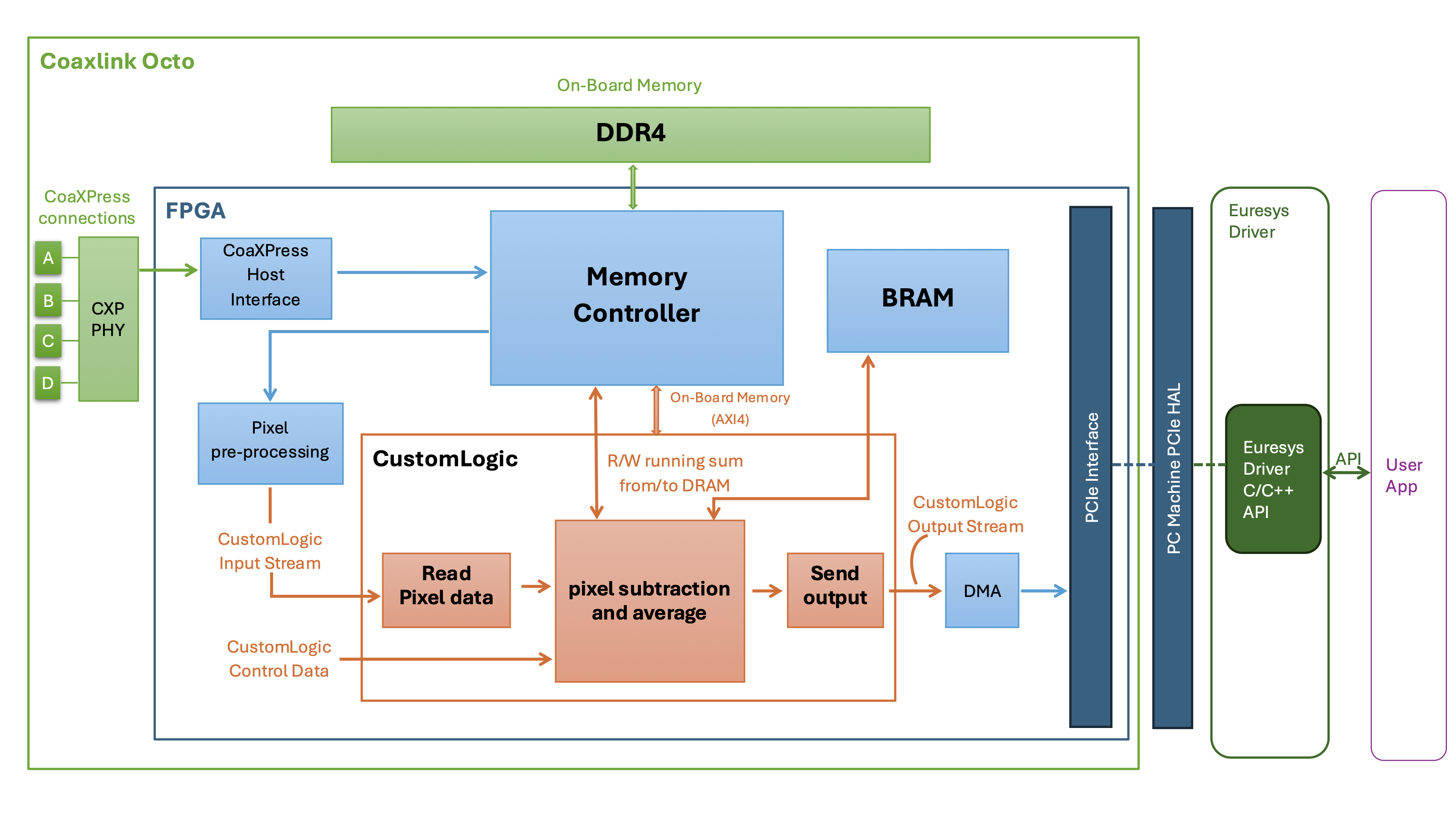}
    \caption{Dataflow diagram for the CustomLogic FPGA module integrated within the Coaxlink Octo frame grabber. Incoming pixel data are streamed through the CoaXPress interface into memory controller and then forwarded to the CustonLogic region. Preprocessing kernels perform real-time pixel subtraction and averaging, with intermediate data buffered in DDR4 via the AXI4 protocol and final results routed to host memory via PCIe.}
    \label{fig:fpga-dataflow}
\end{figure}

The hardware setup for PRISM experiments, as described in~\cite{Wu2025}, utilizes a Coaxlink Octo PCIe frame grabber. Figure~\ref{fig:fpga-dataflow} provides an overview of the dataflow during the data acquisition and preprocessing stages. Each board interfaces with a $4$-connection area scan camera via channels A--D, transmitting image data with a stream width of $128$ bits. The number of pixels per packet depends on the selected pixel depth; for example, with $16$-bit encoding, the board receives $8$ pixels per clock cycle. Incoming data are first routed through the memory controller, then forwarded to the CustomLogic region, where FPGA modules perform real-time preprocessing operations.


\subsection{DRAM Access Patterns and Optimization Techniques}
The PRISM preprocessing logic comprises three modules designed to operate directly on streamed data. First, the \emph{input digestion module} reads incoming packets and forwards them to the \emph{image preprocessing module}. After unpacking the pixel-level content, this kernel performs subtraction and averaging operations on-the-fly. A key design constraint is achieving high-bandwidth access to on-board DRAM via the AXI4-stream interface to the memory controller. Once preprocessing is complete, output frames are repacked to match the stream data width and routed through the \emph{send output module}, which transfers results to host memory via the PCIe interface. The remainder of this subsection presents three designs of the image preprocessing module, each developed to meet high-throughput real-time demands.

\subsection*{Algorithm 1: Sequential Access without Burst Mode}
Algorithm~\ref{alg:sub-avg-1} introduces the first design for subtraction and averaging in the preprocessing pipeline. For each PRISM experiment, this algorithm is invoked $N \times G$ times, as the CustomLogic module is triggered upon receipt of every incoming frame. Image data are transferred in packets rather than full frames, so the CustomLogic module operates concurrently with packet delivery---ideally initiating processing as soon as each packet arrives and completing it before the next one. For clarity, and without loss of generality, we assume each packet contains a single pixel throughout this section, although the true payload may vary depending on pixel depth and hardware configuration.


While the core computations in PRISM preprocessing are conceptually straightforward, a key design challenge lies in managing intermediate frame storage. These frames, computed by subtracting pairs of consecutive input frames, must be retained across groups for subsequent averaging. A natural implementation strategy for Algorithm~\ref{alg:sub-avg-1} would use FPGA Block RAM (BRAM) to store all intermediate results; however, this becomes infeasible at PRISM-scale due to limited BRAM capacity, especially when the number of groups $G$ and frames per group $N$ can be substantial. The Coaxlink Octo board shown in Figure~\ref{fig:fpga-dataflow} offers roughly $12.35$ MB of BRAM for application use, and even high-end FPGAs typically provide only tens of MBs. In contrast, PRISM’s intermediate frame data can easily exceed hundreds of MBs. Accordingly, Algorithm~\ref{alg:sub-avg-1} offloads these intermediate results to on-board DRAM, storing them in the array \texttt{tmpFrame}. A group counter $g$ and frame counter $i$---updated with each kernel invocation---track the storage index for incoming pixels.


While this approach alleviates the storage limitations, it introduces a significant bottleneck: the latency of data transfers between the processing logic and on-board DRAM. In Algorithm~\ref{alg:sub-avg-1}, memory access occurs in two latency-sensitive phases---pixel-by-pixel writes during intermediate frame generation (line~\ref{line:algo1-dram-write}), and pixel-by-pixel reads during final averaging (line~\ref{line:algo1-dram-read}). In total, $(G - 1) \times H \times W\times\frac{N}{2}$ write operations are issued to DRAM to construct the intermediate dataset. Subsequently, the full set of stored frames must be read back to compute groupwise sums. Beyond the sheer volume of reads, performance is further constrained by noncontiguous DRAM access, which inhibits pipelining and degrades memory throughput.


\begin{algorithm}[H]
    \caption{Sequential Access without Burst Mode}
    \label{alg:sub-avg-1}
    \begin{algorithmic}[1]
    \STATE i: frame counter ($1\leq$ i $\leq N$ )
    \STATE g: group counter ($1\leq$ g $\leq G$ )
    \STATE prvFrame[$H\times W$]: previous frame in BRAM
    \STATE outFrame[$H\times W$]: output frames in BRAM
    \STATE tmpFrame[$G$][$N$/2][$H\times W$]: array of intermediate frames in DRAM 
    \STATE 
    \STATE Update i and g
    \STATE 
    \STATE \textbf{for} each pixel j in frame i:
    \STATE \hspace{0.3cm} val $\leftarrow$ read pixel j
    \STATE \hspace{0.3cm} \textbf{if} g $ \neq G$: 
    \STATE \hspace{0.6cm} \textbf{if} odd i:
    \STATE \hspace{0.9cm} prvFrame[j] $\leftarrow$ val
    \STATE \hspace{0.6cm} \textbf{else}:
    \STATE \hspace{0.9cm} tmpFrame[g][i/2][j] $\leftarrow$ val - prvFrame[j] \label{line:algo1-dram-write}
    \STATE \hspace{0.3cm} \textbf{else}:
    \STATE \hspace{0.6cm} \textbf{if} odd i:
    \STATE \hspace{0.9cm} prvFrame[j] $\leftarrow$ val
    \STATE \hspace{0.6cm} \textbf{else}: 
    \STATE \hspace{0.9cm} \textbf{for} each group h from 1 to $G-1$
    \STATE \hspace{1.2cm} outFrame[j] $+=$ tmpFrame[h][i][j] \label{line:algo1-dram-read}
    \STATE \hspace{0.9cm} outFrame[j] $+=$ val - prvFrame[j]
    \STATE \hspace{0.9cm} outFrame[j] $/=G$
    \end{algorithmic}
\end{algorithm}

\subsection*{Algorithm 2: Optimized Write with Burst Mode}
To mitigate these latency issues, we propose Algorithm~\ref{alg:sub-avg-2}, a modified version of Algorithm~\ref{alg:sub-avg-1}. The key idea is to defer DRAM writes until all necessary data are available, enabling the Vitis HLS compiler to fully pipeline the write process and activate burst mode via the AXI4-stream interface. This optimization is designed to reduce write latency by consolidating pixel transfers into high-throughput bursts. Importantly, while timing improves, the total number of pixels written to DRAM remains unchanged.


\begin{algorithm}[H]
    \caption{Optimized Write with Burst Mode}
    \label{alg:sub-avg-2}
    \begin{algorithmic}[1]
    \STATE i: frame counter ($1\leq$ i $\leq N$)
    \STATE g: group counter ($1\leq$ g $\leq G$)
    \STATE prvFrame[$H\times W$]: previous frame in BRAM
    \STATE subFrame[$H\times W$]: frame subtraction result in BRAM
    \STATE outFrame[$H\times W$]: output frames in BRAM
    \STATE tmpFrame[$G$][$N$/2][$H\times W$]: array of intermediate frames in DRAM
    \STATE
    \STATE Update i and g
    \STATE
    \STATE \textbf{for} each pixel j in frame i:
    \STATE \hspace{0.3cm} val $\leftarrow$ read pixel j
    \STATE \hspace{0.3cm} \textbf{if} g $ \neq G$: 
    \STATE \hspace{0.6cm} \textbf{if} odd i:
    \STATE \hspace{0.9cm} prvFrame[j] $\leftarrow$ val
    \STATE \hspace{0.6cm} \textbf{else}:
    \STATE \hspace{0.9cm} subFrame[j] $\leftarrow$ val - prvFrame[j]
    \STATE \hspace{0.3cm} \textbf{else}:
    \STATE \hspace{0.6cm} \textbf{if} odd i:
    \STATE \hspace{0.9cm} prvFrame[j] $\leftarrow$ val
    \STATE \hspace{0.6cm} \textbf{else}: 
    \STATE \hspace{0.9cm} \textbf{for} each group h from 1 to $G-1$
    \STATE \hspace{1.2cm} outFrame[j] $+=$ tmpFrame[h][i][j]
    \STATE \hspace{0.9cm} outFrame[j] $+=$ val - prvFrame[j]
    \STATE \hspace{0.9cm} outFrame[j] $/=G$    
    \STATE
    \STATE \textbf{if} g $\neq G$ and i is even:
    \STATE \hspace{0.3cm} \textbf{for} each pixel j in frame i:
    \STATE \hspace{0.6cm} tmpFrame[g][i/2][j] $\leftarrow$ subFrame[j]
    \STATE
    \end{algorithmic}
\end{algorithm}

\subsection*{Algorithm 3: Optimized Read and Write with Pipelined Accumulation}
\begin{figure}[H]
    \centering
    \includegraphics[scale=0.07]{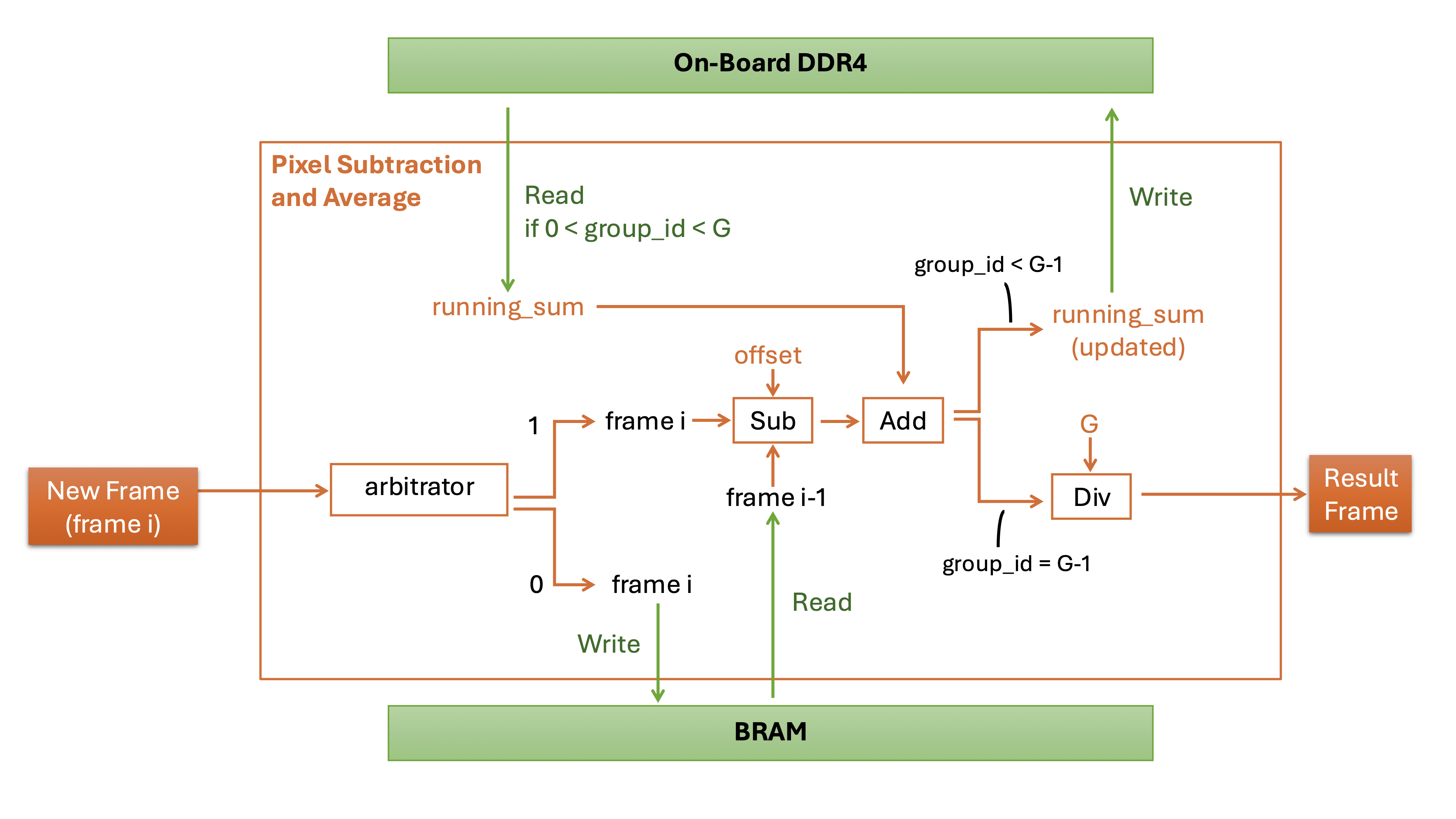}
    \caption{Dataflow illustration for Algorithm~\ref{alg:sub-avg-3}, highlighting optimized burst-mode interactions with on-board DRAM. Incoming image frames undergo pixel-wise subtraction and incremental accumulation, with results written to DRAM as a running sum. Read and write operations are burst-enabled through the AXI4 interface, improving throughput and minimizing latency. The diagram outlines the sequence of memory transactions and accumulation logic across frame groups, culminating in the final averaged output.}
    \label{fig:algo3-flow}
\end{figure}

A key drawback of Algorithm~\ref{alg:sub-avg-2} is that it still performs pixel-by-pixel reads from DRAM, limiting throughput despite write-side optimizations. To further accelerate processing, we extend the burst-mode strategy to DRAM reads---requiring a more fundamental reworking of Algorithm~\ref{alg:sub-avg-1}.The central idea is to incrementally accumulate the groupwise sum as new subtraction data become available, rather than deferring summation until all intermediate frames have been collected. This approach replaces storage of individual subtraction results with a running sum, which is written directly to DRAM. The revised strategy, detailed in Algorithm~\ref{alg:sub-avg-3}, enables burst mode for both DRAM reads and writes. As a result, the total volume of DRAM reads is reduced to just $H \times W \times \frac{N}{2}$ pixels---substantially improving memory efficiency over previous designs. Figure~\ref{fig:algo3-flow} illustrates the full dataflow of Algorithm~\ref{alg:sub-avg-3}, highlighting the pipelined accumulation strategy and burst-mode DRAM interactions.


\begin{algorithm}[H]
    \caption{Optimized Read and Write with Pipelined Accumulation}
    \label{alg:sub-avg-3}
    \begin{algorithmic}[1]
    \STATE i: frame counter ($1\leq$ i $\leq N$)
    \STATE g: group counter ($1\leq$ g $\leq G$)
    \STATE prvFrame[$H\times W$]: previous frame in BRAM
    \STATE sumFrame[$H\times W$]: running sum in BRAM
    \STATE outFrame[$H\times W$]: output frames in BRAM
    \STATE tmpFrame[$G$][$N$/2][$H\times W$]: array of intermediate frames in DRAM
    \STATE
    \STATE \textbf{if} g $\neq 1$ and i is even:
    \STATE \hspace{0.3cm} \textbf{for} each pixel j in frame i:
    \STATE \hspace{0.6cm} sumFrame[j] $\leftarrow$ tmpFrame[g][i/2][j]
    \STATE
    \STATE \textbf{for} each pixel j in frame i:
    \STATE \hspace{0.3cm} val $\leftarrow$ read pixel j
    \STATE \hspace{0.3cm} \textbf{if} g $\neq G$: 
    \STATE \hspace{0.6cm} \textbf{if} odd i:
    \STATE \hspace{0.9cm} prvFrame[j] $\leftarrow$ val
    \STATE \hspace{0.6cm} \textbf{else}:
    \STATE \hspace{0.9cm} sumFrame[j] $+=$ val - prvFrame[j]
    \STATE \hspace{0.3cm} \textbf{else}:
    \STATE \hspace{0.6cm} \textbf{if} odd i:
    \STATE \hspace{0.9cm} prvFrame[j] $\leftarrow$ val
    \STATE \hspace{0.6cm} \textbf{else}:
    \STATE \hspace{0.9cm} sumFrame[j] $+=$ val - prvFrame[j]
    \STATE \hspace{0.9cm} sumFrame[j] $/=G$
    \STATE
    \STATE \textbf{if} g $\neq G$ and i is even:
    \STATE \hspace{0.3cm} \textbf{for} each pixel j in frame i:
    \STATE \hspace{0.6cm} tmpFrame[g][i/2][j] $\leftarrow$ sumFrame[j]
    \STATE
    \STATE Update i and g
    \end{algorithmic}
\end{algorithm}

One limitation of Algorithm~\ref{alg:sub-avg-3} is its vulnerability to data type overflow in the running sum, which constrains the maximum number of groups $G$. For example, if input pixels are encoded using $12$ bits and the accumulation is performed with $16$-bit unsigned integers, overflow occurs when $G > 8$. This stems from the fact that each $12$-bit pixel is internally wrapped to $16$ bits, but the cumulative sum across groups may still exceed the representable range. To resolve this, Algorithm~\ref{alg:sub-avg-3} can be modified to ``spread'' the division operations---dividing each incoming difference by $G$ before it is added to the running sum. This ensures that accumulation remains bounded by the output data type and safely accommodates arbitrarily large values of $G$ without overflow risk.

Before concluding this section, we offer two implementation notes regarding Algorithms~\ref{alg:sub-avg-1} to \ref{alg:sub-avg-3}. First, each algorithm can be modified to return the averaged frames in an additional group---specifically, the $(G+1)$‑th group---interleaved with raw frame data. This optional feature is valuable for verifying algorithmic correctness during development. However, because the original raw data are typically not required in downstream scientific analyses, we instead compute and transmit the average immediately after the final subtraction in group $G$. This more efficient strategy is reflected in Figure~\ref{fig:prism-preprocess}, where averaged frames appear directly in the output of experiment $G$. Second, a fixed offset is added before each subtraction to ensure the resulting values remain within the representable range of the output data type. This offset is subtracted post-transfer to the host to recover original signal amplitudes without loss. In the next section, we describe the experimental setup that simulates actual PRISM acquisition behavior; this configuration is also adopted in the hardware validation presented in Section~\ref{sec:fpga-perform}.



\section{Hardware Setup and Acquisition Emulation}\label{sec:experiment}
\begin{figure}[H]
    \centering
    \includegraphics[scale=0.5]{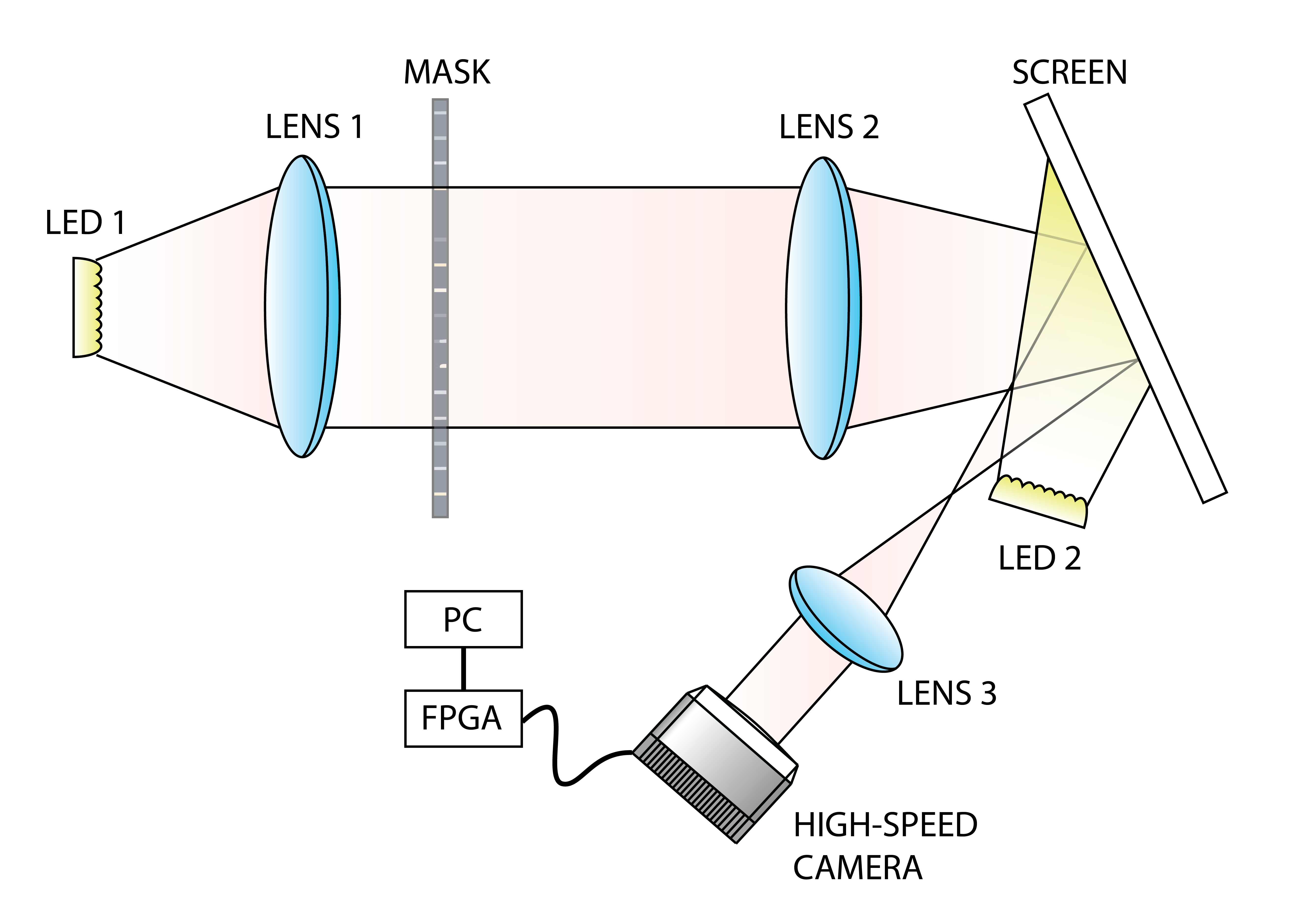}
    \caption{Experimental hardware setup for PRISM emulation, integrating FPGA-based preprocessing within the acquisition pipeline. A Phantom S710 high-speed camera captures dynamic screen patterns illuminated by two LEDs---one modulated to simulate transient excitation, the other static as background noise. Image data are streamed via CoaXPress into the Coaxlink Octo frame grabber, where an embedded Xilinx FPGA performs real-time denoising before host transfer. The configuration mimics excitation-driven workflows and validates low-latency performance under realistic operating conditions.}
    \label{fig:exp-setup}
\end{figure}

To evaluate our framework under realistic conditions, we implemented it on a Coaxlink Octo PCIe frame grabber board featuring an embedded Xilinx Kintex Ultrascale XCKU035 FPGA and $2$ GB of on-board DDR4 memory. Real-time performance was validated using a Phantom S710 high-speed camera, which captures raw image data from a fixed screen pattern. The camera includes $16$ CoaXPress connections arranged in four banks of four channels, enabling high-bandwidth data acquisition. Captured frames are streamed via CoaXPress into the frame grabber for on-board preprocessing. To simulate PRISM’s alternating excitation pattern, two LEDs were positioned in front of the screen: one modulated by a sine-wave signal to mimic transient dynamics, and a second with constant luminance serving as ambient noise. The camera acquires images at a preset frequency, emulating PRISM’s temporal behavior. Figure~\ref{fig:exp-setup} illustrates the complete hardware setup used for this validation. The following section assesses the real-time feasibility and comparative performance of the three FPGA preprocessing algorithms through HLS simulation, benchmarked against this experimental setup.


\section{Latency Benchmarks and Protocol-Aware Analysis}\label{sec:hls-performance}
This section evaluates the performance of Algorithms~\ref{alg:sub-avg-1} to \ref{alg:sub-avg-3} using AMD Vitis High-Level Synthesis (HLS) simulation reports. All three algorithms are implemented in C++, with register-transfer level (RTL) code generated via Vivado HLS 2018.3 (formerly known as Vitis HLS). Input images are assumed to follow the mono12 format---that is, $12$-bit unsigned integers internally wrapped in $16$-bit unsigned containers. All arithmetic operations are performed using $16$-bit unsigned integers. Simulations are conducted using fixed parameters: $G = 8$ groups and $N = 1000$ frames per group. For consistency, each algorithm returns results in the final group. The variant referred to as burst R/W v2 corresponds to the modified version of Algorithm~\ref{alg:sub-avg-3} introduced in Section~\ref{sec:methods}.


\begin{table}[H]
    \centering
    \caption{Estimated latency and initiation interval in clock cycles for the core subtraction-and-averaging kernel across all FPGA algorithms. Burst-mode DRAM access in Algorithm~\ref{alg:sub-avg-3} significantly lowers both minimum and maximum clock cycle counts, enabling sustained throughput below the inter-frame threshold. The modified variant (burst R/W v2) incurs negligible latency overhead due to distributed division operations.}
    \label{tab1:latency}
    \begin{tabular}{ccccc}
        \toprule
        \multirow{2}{*}{Algorithm} & \multicolumn{2}{c}{Latency} & \multicolumn{2}{c}{Interval} \\
        & min & max & min & max \\
        \midrule
        No burst mode & $17944$  &  $17944$  &  $17944$  &  $17944$  \\
        burst write & $17938$  &  $20515$  &  $17938$  &  $20515$ \\
        burst R/W & $2570$  &  $7721$  &  $2570$  &  $7721$ \\
        burst R/W v2 & $2575$  &  $7726$  &  $2575$  &  $7726$ \\
        \bottomrule
    \end{tabular}
\end{table}

Table~\ref{tab1:latency} summarizes the estimated latency and initiation intervals for the core subtraction-and-averaging kernel across all three algorithms. Burst-mode DRAM writes in Algorithm~\ref{alg:sub-avg-2} reduce the minimum clock cycle count slightly, though overall gains are limited. This is because DRAM writes occur consistently during the first $G$ groups, with each iteration writing a fixed amount of data. In contrast, DRAM reads begin only during processing of the second group and peak during final averaging, where a total of $G \times H \times W\times\frac{N}{2}$ pixels must be retrieved. As a result, latency remains high. With burst-mode enabled for both reads and writes---as implemented in Algorithm~\ref{alg:sub-avg-3}---the minimum and maximum clock cycle counts drop significantly due to improved throughput. This optimization relies on the fact that only $H\times W\times\frac{N}{2}$ pixels are read during the averaging stage. Finally, the modified version of Algorithm~\ref{alg:sub-avg-3}(labeled as burst R/W v2) introduces a negligible increase in latency due to performing division during each partial sum, slightly increasing the number of arithmetic operations.


\begin{table}[H]
    \small
    \centering
    \caption{Loop-level latency estimates across three FPGA architectures. The burst R/W variants of Algorithm~\ref{alg:sub-avg-3} sustain low latency with consistent initiation intervals, while the other designs incur extra cycles due to non-optimized DRAM access,  resulting in inter-frame stalls.}
    \label{tab2}
    \begin{tabular}{cccccccc}
        \toprule
        Algorithm & \multirow{2}{*}{\shortstack{Loop\\ Name}} & \multirow{2}{*}{Latency} & \multirow{2}{*}{\shortstack{Iteration\\ Latency}} & \multicolumn{2}{c}{Initiation Interval} & \multirow{2}{*}{\shortstack{Trip\\ Count}} & \multirow{2}{*}{Pipelined} \\
         \cline{5-6}
         & & & & achieved & target \\
        \midrule
        No burst mode & \texttt{PixSubAvgLoop}  &  $17935$  &  $23$  &  $7$  &  $1$  &  $2560$  & yes \\
        \midrule
        \multirow{2}{*}{burst write} & \texttt{PixSubAvgLoop}  &  $17934$  &  $22$  &  $7$  &  $1$ &  $2560$  & yes \\
         & \texttt{WriteToDRAMLoop}  &  $2562$  &  $4$  &  $1$  &  $1$ &  $2560$  & yes \\
        \midrule
        \multirow{3}{*}{burst R/W} & \texttt{ReadFromDRAMLoop}  &  $2561$  &  $3$  &  $1$  &  $1$ &  $2560$  & yes \\
         & \texttt{PixSubAvgLoop}  &  $2565$  &  $7$  &  $1$  &  $1$ &  $2560$  & yes \\
         & \texttt{WriteToDRAMLoop}  &  $2562$  &  $4$  &  $1$  &  $1$ &  $2560$  & yes \\
         \midrule
         \multirow{3}{*}{burst R/W v2} & \texttt{ReadFromDRAMLoop}  &  $2561$  &  $3$  &  $1$  &  $1$ &  $2560$  & yes \\
         & \texttt{PixSubAvgLoop}  &  $2570$  &  $12$  &  $1$  &  $1$ &  $2560$  & yes \\
         & \texttt{WriteToDRAMLoop}  &  $2562$  &  $4$  &  $1$  &  $1$ &  $2560$  & yes \\
        \bottomrule
    \end{tabular}
\end{table}

Table~\ref{tab2} presents a breakdown of estimated latency for key processing loops across all algorithms. The \texttt{PixSubAvgLoop} performs the core subtraction and averaging, while \texttt{WriteToDRAMLoop} and \texttt{ReadFromDRAMLoop} handle burst-mode transfers to and from DRAM. Notably, Algorithm~\ref{alg:sub-avg-3} (burst R/W) achieves near-uniform latency distribution across loops, enabling improved throughput via pipelining and loop unrolling where applicable. This architectural balance accounts for its superior performance relative to Algorithms~\ref{alg:sub-avg-1} and \ref{alg:sub-avg-2}, where skewed latency profiles and limited pipelining hinder efficiency. Additionally, spreading division operations---as in the modified version of Algorithm~\ref{alg:sub-avg-3} (burst R/W v2)---introduces minimal performance overhead. The initiation interval for \texttt{PixSubAvgLoop} in Algorithms~\ref{alg:sub-avg-1} and \ref{alg:sub-avg-2} remains at $7$ clock cycles, constraining their ability to sustain frame-by-frame throughput and ultimately preventing real-time operation.


It's important to note that Tables~\ref{tab1:latency} and \ref{tab2} report only loop-level clock cycles, treating each iteration as approximately one cycle. These estimates exclude AXI4 protocol effects, particularly the impact of handshaking and burst-mode behavior. As a result, the latency for Algorithm~\ref{alg:sub-avg-2} appears inflated---the benefits of burst-mode write optimization are not fully captured. Moreover, these tables do not reflect the acceleration enabled by burst-mode reads in Algorithm~\ref{alg:sub-avg-3}. Despite this abstraction, scheduling analysis reveals that Algorithm~\ref{alg:sub-avg-3} consistently maintains sufficient slack between input frames, ensuring uninterrupted processing and real-time performance. Finally, the latency spread shown in Table~\ref{tab1:latency} stems from conditional branches within the pixel subtraction and averaging module.


\begin{figure}[H]
    \centering
    \begin{subfigure}[b]{0.4\textwidth}
        \centering
        \includegraphics[width=0.9\linewidth]{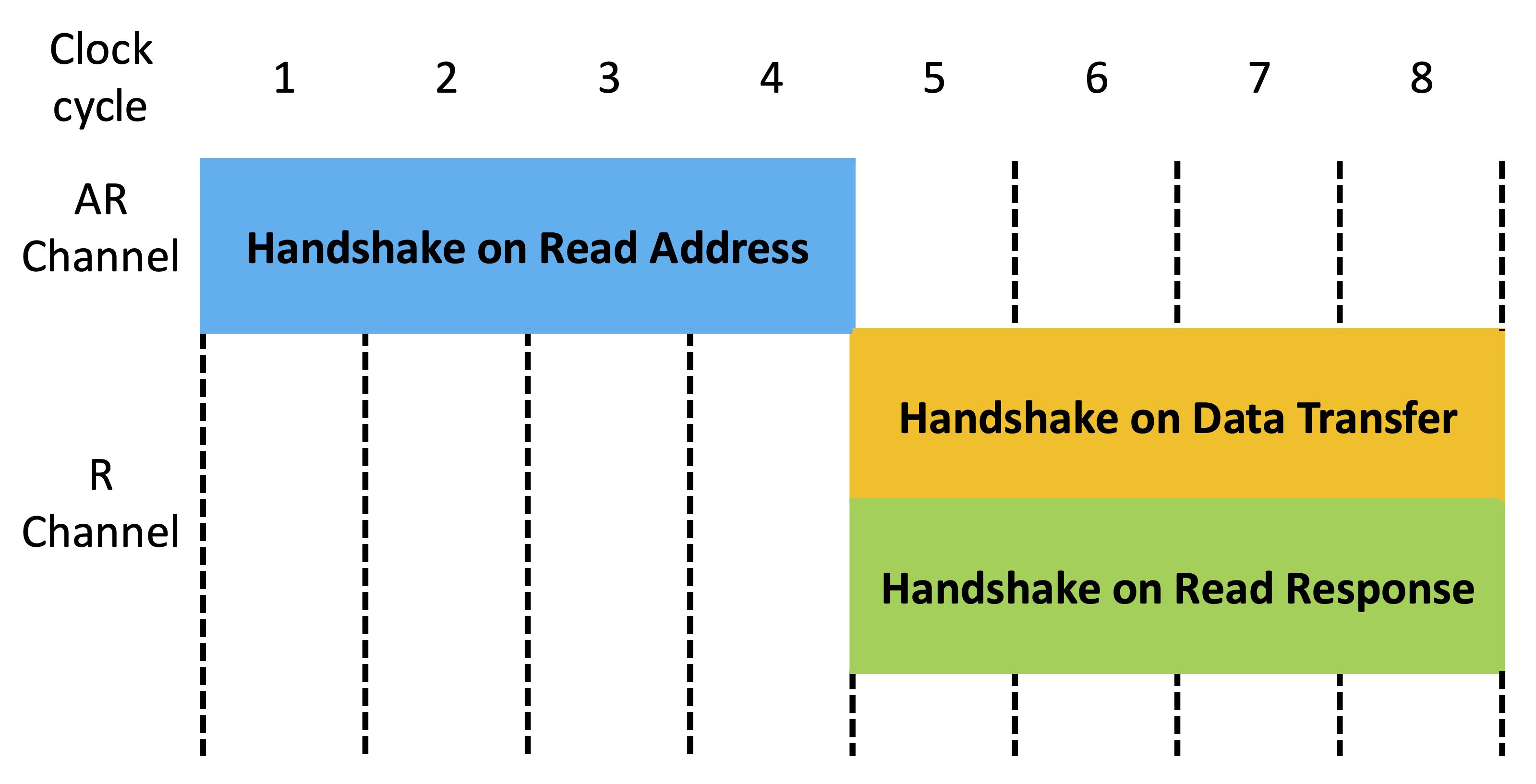}
        \caption{Single-data read}
        \label{fig:AXI4-single-read}
    \end{subfigure}
    \begin{subfigure}[b]{0.4\textwidth}
        \centering
        \includegraphics[width=0.9\linewidth]{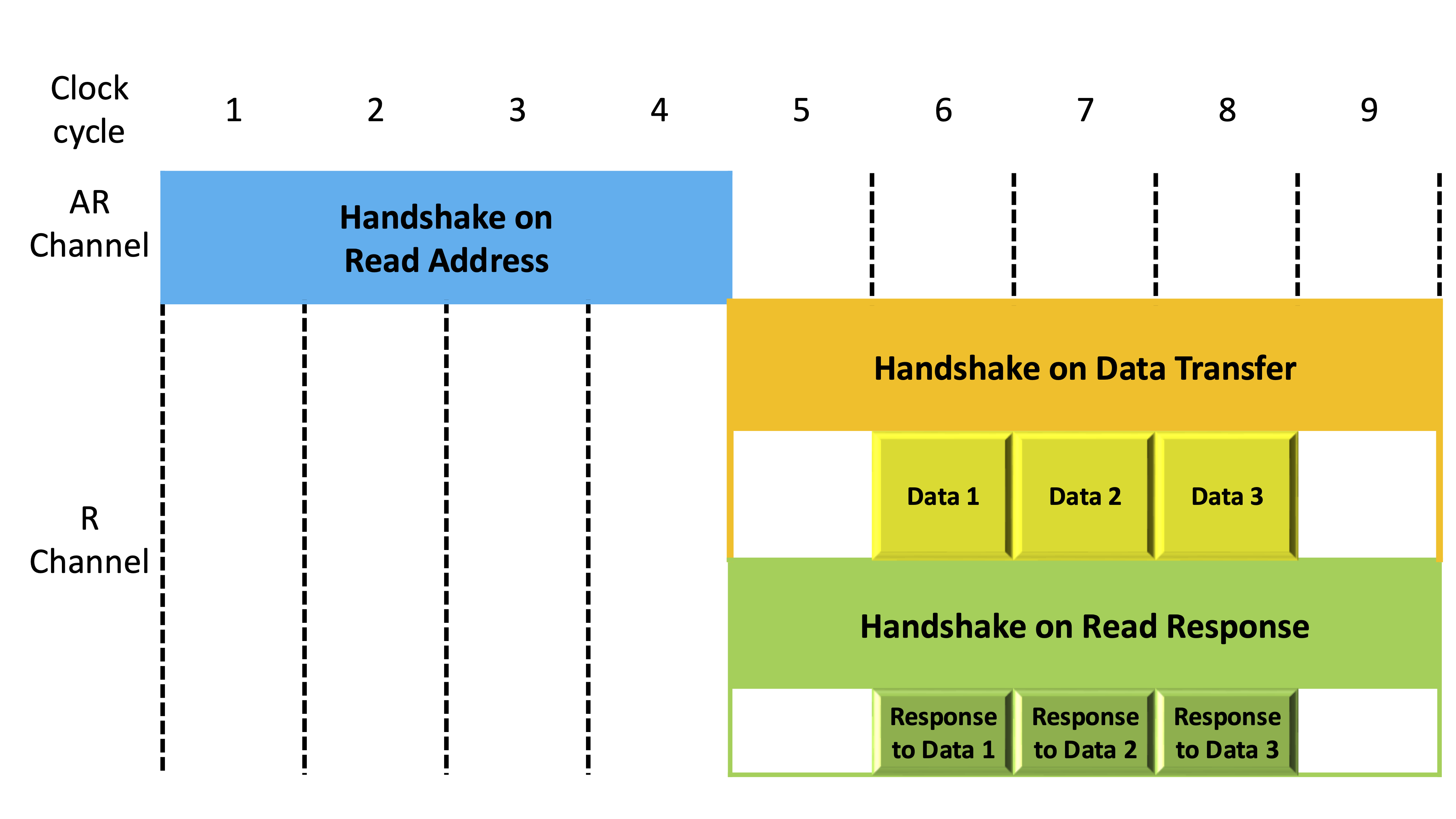}
        \caption{Burst-mode read}
        \label{fig:AXI4-multiple-read}
    \end{subfigure}
    \newline
    \begin{subfigure}[b]{0.4\textwidth}
        \centering
        \includegraphics[width=0.9\linewidth]{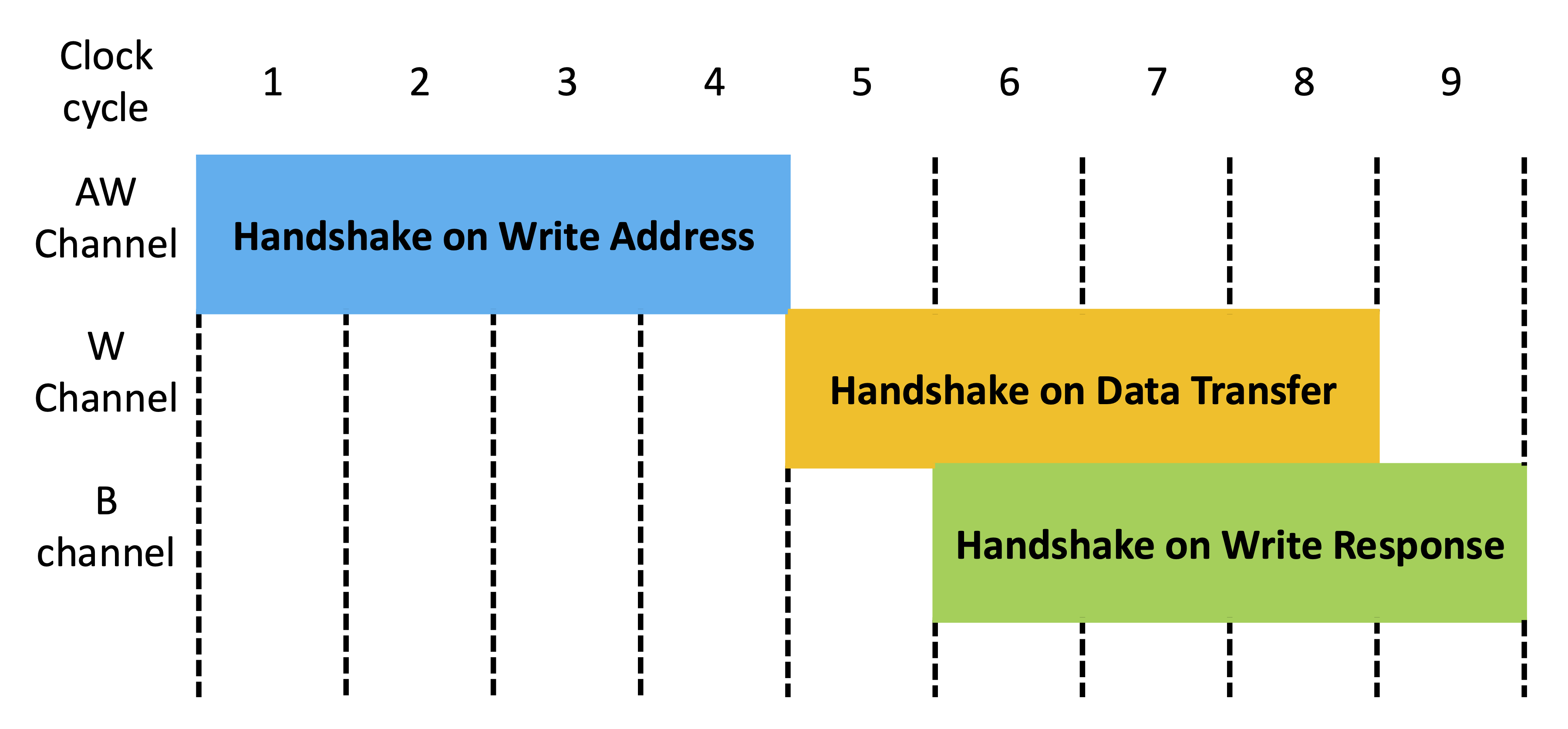}
        \caption{Single-data write}
        \label{fig:AXI4-single-write}
    \end{subfigure}
    \begin{subfigure}[b]{0.4\textwidth}
        \centering
        \includegraphics[width=0.9\linewidth]{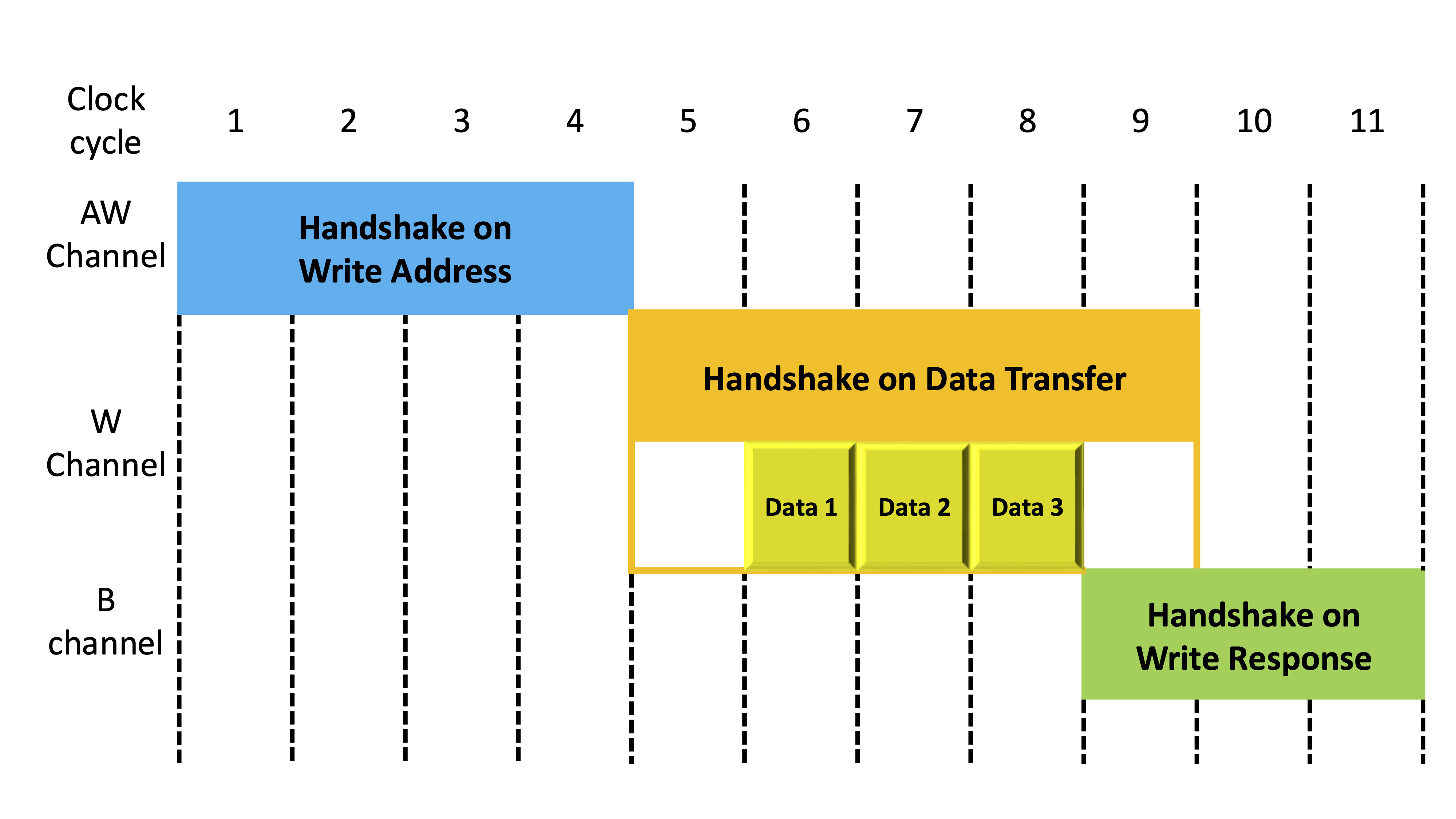}
        \caption{Burst-mode write}
        \label{fig:AXI4-multiple-write}
    \end{subfigure}
    \caption{Read/write mechanism of AXI4 protocol. Timing diagrams of AXI4 protocol for memory transactions. Panels (a)--(d) illustrate the sequence and latency overhead associated with single-data and burst-mode transfers for both read and write operations. In each case, the AXI4 protocol requires a series of handshakes across address, data, and response channels, which significantly influences throughput and initiation intervals in FPGA-based pipelines.}
    \label{fig:AXI4-mechanism}
\end{figure}

A more precise latency estimate can be derived by incorporating AXI4 protocol characteristics, as illustrated in Figure~\ref{fig:AXI4-mechanism}. Single data transfers incur approximately $8$ clock cycles for reads (Figure~\ref{fig:AXI4-single-read}) and $9$ clock cycles for writes (Figure~\ref{fig:AXI4-single-write}). In contrast, burst-mode operations---shown in Figure~\ref{fig:AXI4-multiple-read} and \ref{fig:AXI4-multiple-write}---require about $9$ clock cycles to read and $11$ clock cycles to write three consecutive data packets. Using these timing profiles, and assuming that non-DRAM operations within the \texttt{PixSubAvgLoop} consume $1$ clock cycle per iteration, we estimate algorithmic latencies with a $2$~ns FPGA clock. For odd-numbered frames, which bypass DRAM access in all three designs, the estimated latency is simply:

\[
\frac{2560\times 2}{1000} = 5.12~(\mu s).
\]

\begin{figure}[H]
    \centering
    \includegraphics[width=1.0\linewidth]{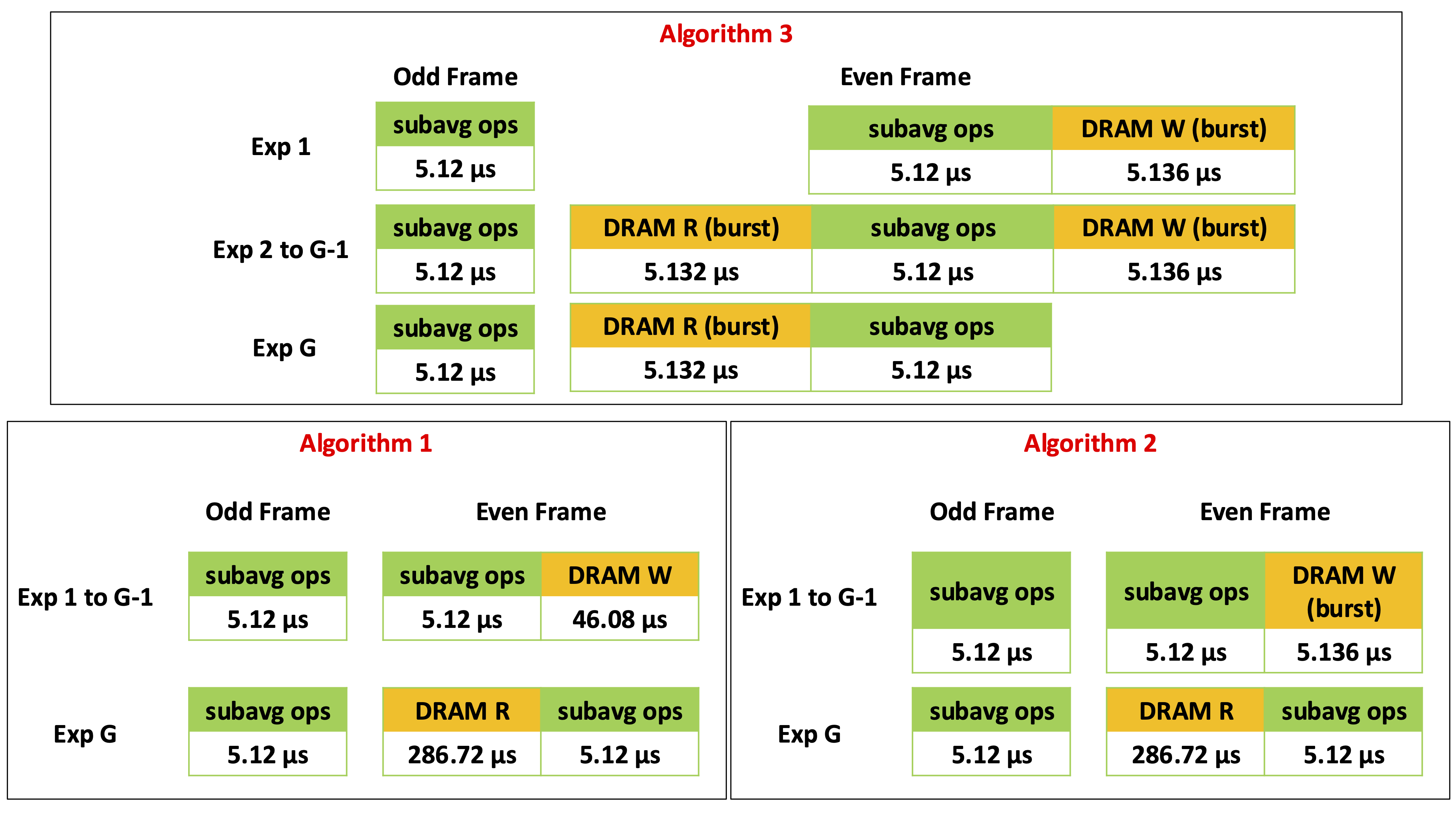}
    \caption{Estimated per-frame latency for three FPGA preprocessing algorithms under PRISM-scale conditions PRISM-scale acquisition. This diagram compares timing contributions from subtraction, averaging, and DRAM access across odd and even frames. Timing estimates assume a $2$~ns FPGA clock and data packet width of $128$ bits. Algorithm~\ref{alg:sub-avg-1} lacks burst-mode optimizations, incurring excessive read/write latency and violating the $57~\mu s$ inter-frame constraint. Algorithm~\ref{alg:sub-avg-2} applies burst-mode writes, improving throughput but leaving read-side delays unaddressed. Algorithm~\ref{alg:sub-avg-3} uses burst-mode reads and writes with pipelined accumulation, sustaining real-time execution.}
    \label{fig:algo-time-estimates}
\end{figure}

Next, consider the even-numbered frames. In all three algorithms, write operations occur during the first $G-1$ groups, while Algorithms~\ref{alg:sub-avg-1} and \ref{alg:sub-avg-2} defer read operations until the final group. Algorithm~\ref{alg:sub-avg-1} does not activate burst mode, so both reads and writes are performed as single data transactions. Accordingly, the estimated processing time per frame for the first $G-1$ groups is
\[
5.12 + \frac{2560\times 9 \times 2}{1000} = 51.2~(\mu s),
\]
whereas the last group requires
\[
\frac{2560\times 7\times 8\times 2}{1000} + 5.12 = 291.84~(\mu s).
\]
These values show that Algorithm~\ref{alg:sub-avg-1} exceeds the inter-frame interval ($57~\mu s$), preventing real-time operation. In Algorithm~\ref{alg:sub-avg-2}, burst mode optimizes DRAM writes, reducing the estimated processing time in the first $G-1$ groups to
\[
5.12 + \frac{(2560+2+4+2) \times 2}{1000} = 10.256~(\mu s).
\]
Although more efficient than Algorithm~\ref{alg:sub-avg-1}, its reliance on non-pipelined reads in the final group still prevents real-time execution. By contrast, Algorithm~\ref{alg:sub-avg-3} activates burst mode for both reads and writes and replaces explicit intermediate frame storage with a running sum that is incrementally updated. In the first group, only a burst-mode write is performed, so its estimated processing time matches that of the first $G-1$ groups in Algorithm~\ref{alg:sub-avg-2}. For the middle groups ($2\leq g < G$), where both burst-mode reads and writes are issued, the estimated latency per frame is
\[
\frac{(2560+4+2)\times 2}{1000} + 5.12 + \frac{(2560+2+4+2) \times 2}{1000} = 15.388~(\mu s).
\]
In the final group, only a burst-mode read is performed, yielding a reduced latency of
\[
\frac{(2560+4+2)\times 2}{1000} + 5.12 = 10.252 \ (\mu s).
\]
These estimates confirm that Algorithm~\ref{alg:sub-avg-3} satisfies the inter-frame interval, enabling sustained real-time denoising in PRISM-scale imaging pipelines. For a visual summary of these performance estimates, see Figure~\ref{fig:algo-time-estimates}.

Using these latency estimates, we calculate the estimated total time $\bar{t}_1$ required to process $8000$ frames for Algorithm~\ref{alg:sub-avg-1}, with parameters $N=1000$ and $G=8$. Assuming the camera generates one frame every $57~\mu s$, the actual per-frame execution time is approximated as the maximum between the computational latency and the inter-frame arrival rate. The resulting expression is
\[\begin{aligned}
\bar{t}_{1} &= \max(5.12, 57)\times 4000 + \max(51.2,57)\times 3500 + 291.84\times 500 \\
&= 573420~(\mu s) \\
&= 0.5734~(s)
\end{aligned}\]
Compared to the hardware-measured value from Table~\ref{tab3}, $t_1=2.244~s$, the discrepancy is
\[
t_1-\bar{t}_1 = 2.244 - 0.5734 = 1.6706~(s).
\]
This gap suggests that the actual initiation interval is significantly higher than what is inferred from HLS simulation. Approximating the effective initiation interval yields
\[
\frac{1.6706\times 10^9}{2\times 8000\times 2559}\approx 41~(\text{clock cycles}).
\]
This analysis highlights that Table~\ref{tab2} underestimates the required initiation interval. A key contributor to this limitation is the hardware implementation’s use of a single AXI4 port, which processes only one read or write request at a time, creating a serialization bottleneck that is not captured by simulation. 

For Algorithm~\ref{alg:sub-avg-2}, we estimate the total processing time across $8000$ frames as follows. Using the same camera frame interval of $57~\mu s$, the per-frame latency is again taken as the maximum between computational time and frame arrival rate. The projected execution time becomes
\[\begin{aligned}
\bar{t}_{2} &= \max(5.12, 57)\times 4000 + \max(10.256, 57)\times 3500 + 291.84\times 500 \\
&= 573420~(\mu s) \\
&= 0.5734~(s),
\end{aligned}\]
Compared with the measured hardware execution time from Table~\ref{tab3}, $t_2=1.092~s$, the timing gap is
\[
t_2-\bar{t}_2 = 1.092 - 0.5734 = 0.5186~(s).
\]
From this, the effective initiation interval on hardware can be estimated as
\[
\frac{0.5186\times 10^9}{2\times 8000\times 2559}\approx 13~(\text{clock cycles}).
\]
This analysis confirms that enabling burst mode for DRAM write operations alone, as in Algorithm~\ref{alg:sub-avg-2}, substantially lowers the effective initiation interval compared to Algorithm~\ref{alg:sub-avg-1}, leading to improved throughput and partial acceleration---although full real-time performance still remains out of reach due to the absence of read-side optimizations.

For Algorithm~\ref{alg:sub-avg-3}, which applies burst-mode optimization to both DRAM reads and writes, the estimated execution time across $8000$ frames is
\[\begin{aligned}
\bar{t}_{3} &= \max(5.12, 57)\times 4000 + \max(10.256, 57)\times 500\\ & \quad + \max(15.388, 57)\times 3000 + \max(10.252, 57)\times 500 \\
&= 456000~(\mu s) \\
&= 0.456~(s),
\end{aligned}\]
This estimate closely matches the measured runtime $t_3=0.457~s$. indicating that the kernel sustains processing within the camera's inter-frame interval. Hence, we infer that the design maintains continuous data throughput with essentially no pipeline stalls, and the initiation interval is approximated as its possible minimum value $1$ clock cycle. This outcome demonstrates that Algorithm~\ref{alg:sub-avg-3} meets the real-time performance threshold, achieving minimal latency and maximizing FPGA efficiency. It significantly outperforms the preceding algorithms by aligning frame-by-frame processing with memory bandwidth and protocol constraints. The next section assesses real-world performance of the three FPGA preprocessing algorithms, contrasts Algorithm~\ref{alg:sub-avg-3} with CPU and GPU workflows, and verifies the kernel’s denoising effectiveness.

\section{Real-Time FPGA Execution and Scalability}\label{sec:fpga-perform}

To evaluate the performance of the three FPGA preprocessing algorithms, we conduct two sets of experiments. The first uses software-triggered acquisition, where the Coaxlink board’s internal driver controls the trigger signal. The second uses an external LED source with frequency configured at $5$~kHz to emulate excitation-driven triggering. In both cases, experiments are performed using $G=8$ groups of $N=1000$ frames, producing $500$ results per test that are returned within the final group. We compare the average elapsed time for generating these results, the effective frame rate, and the corresponding data throughput. Summary statistics for each implementation are presented in Table~\ref{tab3} and Table~\ref{tab4}.


In these experiments, the camera’s minimum cycle period is set to $57~\mu s$. The results demonstrate that our proposed algorithm---Algorithm~\ref{alg:sub-avg-3}---achieves real-time data processing on the Coaxlink Octo board. By dividing the average elapsed time by the total number of samples ($N\times G$), the per-frame latency for Algorithm~\ref{alg:sub-avg-3} is approximately $57.12~\mu s$, closely matching the camera’s cycle period and confirming sustained real-time throughput. Notably, the burst-mode write variant also outperforms the non-burst implementation in hardware tests, as it better exploits the AXI4 protocol for efficient DRAM transfers.


\begin{table}[H]
    \centering
    \caption{Performance comparison of FPGA preprocessing algorithms under software-triggered acquisition. Average elapsed time, frame rate, and data throughput are reported using $256\times 80$-pixel frames for each implementation. Algorithm~\ref{alg:sub-avg-3} (burst read/write) achieves sustained real-time throughput, closely matching the camera’s inter-frame interval of $57~\mu s$. In contrast, Algorithms~\ref{alg:sub-avg-1} and \ref{alg:sub-avg-2} fall short due to non-pipelined DRAM access.}
    \label{tab3}
    \begin{tabular}{cccc}
        \toprule
        Implementation & \multirow{2}{*}{\shortstack{Avg Elapsed\\ Time (sec.)}} & \multirow{2}{*}{\shortstack{Frame Rate\\ (fps)}} & \multirow{2}{*}{\shortstack{Data Rate\\ (MB/s)}} \\
        & & & \\
        \midrule
        No Burst Mode  &  $2.244$  &  $2901\sim 4153$  &  $119\sim 170$ \\
        W in Burst Mode  &  $1.092$  &  $7315\sim 7511$  &  $300\sim 308$ \\
        R/W in Burst Mode  &  $0.457$  &  $17544$  &  $719$ \\
        \bottomrule
    \end{tabular}
\end{table}

\begin{table}[H]
    \centering
    \caption{Performance comparison of FPGA preprocessing algorithms under LED-triggered acquisition. Algorithm~\ref{alg:sub-avg-3} again achieves near-camera-throughput performance, while Algorithms~\ref{alg:sub-avg-1} and \ref{alg:sub-avg-2} remain bottlenecked by sequential memory transactions. Metrics include elapsed time, frame rate, and sustained throughput using $256\times 80$-pixel frames.}
    \label{tab4}
    \begin{tabular}{cccc}
        \toprule
        Implementation & \multirow{2}{*}{\shortstack{Avg Elapsed\\ Time (sec.)}} & \multirow{2}{*}{\shortstack{Frame Rate\\ (fps)}} & \multirow{2}{*}{\shortstack{Data Rate\\ (MB/s)}} \\
        & & & \\
        \midrule
        No Burst Mode  &  $2.214$  &  $2969\sim 4157$  &  $122\sim 170$ \\
        W in Burst Mode  &  $1.930$  &  $3968\sim 5000$  &  $163\sim 205$ \\
        R/W in Burst Mode  &  $1.601$  &  $5000$  &  $205$ \\
        \bottomrule
    \end{tabular}
\end{table}



Table~\ref{tab5} reports the performance of the R/W burst-mode implementation under varying trigger modes and data sizes. In these experiments, each data bank contains $256\times 80$ pixels. For the two-bank configuration, two FPGA cards are used, with each card processing a single data bank independently. The results indicate that Algorithm~\ref{alg:sub-avg-3} is scalable when the input stream is partitioned across multiple banks and distributed across separate processing units. Notably, the average elapsed time, frame rate, and data throughput remain consistent when scaling from one bank to two, confirming that the architecture supports parallel extension without degradation.


\begin{table}[H]
    \small
    \centering
    \caption{Scalability of R/W burst-mode FPGA preprocessing under variable trigger modes and input dimensions. Average elapsed time, frame rate, and data throughput are reported for both single-bank ($256\times 80$) and dual-bank ($256\times 160$) input streams. Results demonstrate consistent performance across LED-triggered and software-triggered acquisition modes. Parallel execution across two FPGA boards confirms pipeline scalability without added latency.}
    \label{tab5}
    \begin{tabular}{ccccc}
        \toprule
        Trigger Mode & Data Size & \multirow{2}{*}{\shortstack{Avg Elapsed\\ Time (sec.)}} & \multirow{2}{*}{\shortstack{Frame Rate\\ (fps)}} & \multirow{2}{*}{\shortstack{Data Rate\\ (MB/s)}} \\
        & & & & \\
        \midrule
        \multirow{2}{*}{Software} & $1$ bank ($256\times 80$) &  $0.457$  &  $17544$  &  $719$ \\
         & $2$ banks ($256\times 160$) &  $0.457$  &  $17544$  &  $719$ \\
        \multirow{2}{*}{LED} & $1$ bank ($256\times 80$) &  $1.601$  &  $5000$  &  $205$ \\
         & $2$ banks ($256\times 160$) &  $1.601$  &  $5000$  &  $205$ \\
        \bottomrule
    \end{tabular}
\end{table}

Table~\ref{tab6} presents the average elapsed time of Algorithm~\ref{alg:sub-avg-3} under varying group counts, with each experiment configured to process $1000$ frames per group at the maximum frame rate supported by the Octo board. The results confirm that the FPGA preprocessing kernel maintains consistent per-frame latency---even as the number of groups increases. This underscores the algorithm’s scalability and its ability to accommodate extended acquisition sequences without degrading real-time performance.


\begin{table}[H]
    \centering
    \caption{Latency consistency of FPGA preprocessing using burst-mode R/W (v2) under varying group counts and input dimensions. Each experiment processes $1000$ frames per group using software-triggered acquisition. Results show stable per-frame latency ($\sim 57~\mu s$) across both single-bank ($256\times 80$) and dual-bank ($256\times 160$) configurations, confirming the scalability of the pipeline. Increasing the number of groups does not degrade throughput, demonstrating sustained real-time performance regardless of sequence depth.}
    \label{tab6}
    \begin{tabular}{cccc}
        \toprule
        \multirow{2}{*}{Data Size} & \multirow{2}{*}{\shortstack{Number of\\ Groups}} & \multirow{2}{*}{\shortstack{Avg Elapsed\\ Time (sec.)}} & \multirow{2}{*}{\shortstack{Elapsed time\\ per frame ($\mu s$)}} \\
        & & & \\
        \midrule
        \multirow{3}{*}{\shortstack{$1$ bank\\ ($256\times 80$)}} & $5$ &  $0.287$ & $57.40$ \\
         & $8$ &  $0.457$ & $57.12$ \\
         & $10$ & $0.571$ & $57.10$ \\
        \multirow{3}{*}{\shortstack{$2$ banks\\ ($256\times 160$)}} & $5$  &  $0.286$ & $57.20$ \\
         & $8$ &  $0.457$ & $57.12$ \\
         & $10$ & $0.571$ & $57.10$ \\
        \bottomrule
    \end{tabular}
\end{table}

To benchmark against alternative workflows, we first evaluated a CPU-based approach using the same data dimensions per bank as in previous FPGA tests. This method buffers raw data from the board directly into host memory before parallel processing via CPU threads---effectively eliminating disk I/O bottlenecks. As shown in Table~\ref{tab7}, increasing thread count improves performance, with the $64$-thread configuration reducing total processing time to approximately $1.05$~s for one bank and $2.06$~s for two banks. However, the approach remains slower than the FPGA implementation, primarily due to unavoidable data transfer overheads between the board and host, which become more pronounced as throughput scales.


\begin{table}[H]
    \centering
    \caption{CPU processing time across varying thread counts for data buffered from FPGA hardware. Tests use one and two data banks ($256\times 80$ and $256\times 160$), with threads applied after host-side buffering. While increased parallelism improves throughput, even $64$-thread execution remains slower than FPGA burst-mode denoising due to data transfer overheads and limited concurrency.}
    \label{tab7}
    \begin{tabular}{ccc}
        \toprule
        \multirow{2}{*}{Number of Threads} & \multicolumn{2}{c}{Avg Total Elapsed Time (sec.)} \\
        \cline{2-3}
        & 1 Bank ($256\times 80$) & 2 Banks ($256\times 160$) \\
        \midrule
        $1$ (sequential) & $34.103$ & $70.753$ \\
        $2$ & $17.658$ & $36.533$ \\
        $4$ & $9.207$ & $18.820$ \\
        $8$ & $5.097$ & $10.320$ \\
        $16$ & $2.812$ & $5.795$ \\
        $32$ & $1.644$ & $3.654$ \\
        $64$ & $1.049$ & $2.061$ \\
        \bottomrule
    \end{tabular}
\end{table}



We next evaluated a GPU-based workflow in which image data are read from disk and processed on a single NVIDIA GPU---either V100 or RTX 2080 Ti. As shown in Tables~\ref{tab8} and table~\ref{tab9}, data transfer overhead between host and device significantly contributes to total elapsed time, compounding alongside disk I/O latency. These effects become more pronounced when processing two banks of data: under this configuration, the GPU's total processing time approaches that of the FPGA. However, given the scaling behavior of transfer bottlenecks, this workflow is likely to exceed FPGA latency as input size or parallelism increases.



\begin{table}[H]
    \small
    \centering
    \caption{Post-acquisition latency breakdown for data processing on a single NVIDIA V100 GPU. Elapsed times include disk I/O, host-to-device data transfer, GPU computation, and total end-to-end duration. Compared to inline FPGA processing, the workflow incurs substantial overhead despite efficient GPU execution.}
    \label{tab8}
    \begin{tabular}{ccccc}
        \toprule
        Data Size & \multirow{2}{*}{\shortstack{Avg Disk Reading\\ Time (sec.)}} & \multirow{2}{*}{\shortstack{Avg Data Transfer\\ Time (sec.)}} & \multirow{2}{*}{\shortstack{Avg Total GPU\\ Time (sec.)}} & \multirow{2}{*}{\shortstack{Avg Total Elapsed\\ Time (sec.)}} \\
        & & & & \\
        \midrule
        $1$ bank ($256\times 80$) &  $0.121$ & $0.115$ & $0.118$ &  $0.238$ \\
        $2$ banks ($256\times 160$) & $0.235$ & $0.240$ & $0.243$ &  $0.478$ \\
        \bottomrule
    \end{tabular}
\end{table}

\begin{table}[H]
    \small
    \centering
    \caption{Post-acquisition latency breakdown for data processing on a single NVIDIA RTX 2080 Ti GPU. Latency trends mirror those of the V100, with slightly higher disk and transfer times. End-to-end performance remains bottlenecked by host-level overhead, confirming the advantage of FPGA inline denoising.}
    \label{tab9}
    \begin{tabular}{ccccc}
        \toprule
        Data Size & \multirow{2}{*}{\shortstack{Avg Disk Reading\\ Time (sec.)}} & \multirow{2}{*}{\shortstack{Avg Data Transfer\\ Time (sec.)}} & \multirow{2}{*}{\shortstack{Avg Total GPU\\ Time (sec.)}} & \multirow{2}{*}{\shortstack{Avg Total Elapsed\\ Time (sec.)}} \\
        & & & & \\
        \midrule
        $1$ bank ($256\times 80$) &  $0.135$ & $0.121$ & $0.123$ &  $0.258$ \\
        $2$ banks ($256\times 160$) & $0.270$ & $0.254$ & $0.258$ &  $0.528$ \\
        \bottomrule
    \end{tabular}
\end{table}

\begin{table}[H]
    \scriptsize
    \centering
    \caption{Comparative execution time of FPGA, CPU, and GPU workflows for denoising high-throughput image data. Buffering time includes data transfer from acquisition hardware to host or disk; total time reflects full preprocessing duration. FPGA burst-mode processing executes denoising inline during acquisition, outperforming CPU and GPU workflows where buffering alone matches or exceeds FPGA's total runtime.}
    \label{tab10}
    \begin{tabular}{ccccc}
        \toprule
        \multirow{2}{*}{Workflow} & \multicolumn{2}{c}{Buffering Data Time (sec.)} & \multicolumn{2}{c}{Avg Total Elapsed Time (sec.)} \\
         & 1 bank ($256\times 80$) & $2$ banks ($256\times 160$) & 1 bank ($256\times 80$) & $2$ banks ($256\times 160$) \\
        \midrule
        FPGA R/W in Burst Mode & $-$ & $-$ & $0.457$ & $0.4565$ \\
        CPU 64 threads (buffering) & \multirow{3}{*}{$0.458$} & \multirow{3}{*}{$0.456$} & $1.049$ & $2.185$ \\
        V100 GPU & & & $0.238$ & $0.478$ \\
        RTX 2080 Ti GPU & & & $0.258$ & $0.528$ \\
        \bottomrule
    \end{tabular}
\end{table}

Table~\ref{tab10} consolidates average elapsed times across the tested workflows, highlighting the relative efficiency of the burst-mode FPGA pipeline. Unlike other methods, which require a fixed buffering step prior to processing, our FPGA implementation performs denoising inline during acquisition, eliminating this overhead entirely. Notably, the time required for data buffering in CPU and GPU workflows closely matches the total elapsed time of the FPGA pipeline---emphasizing the advantage of integrating preprocessing directly into the acquisition stage.

\begin{figure}[H]
    \centering
    \begin{subfigure}[b]{\textwidth}
        \centering
        \includegraphics[scale=0.23]{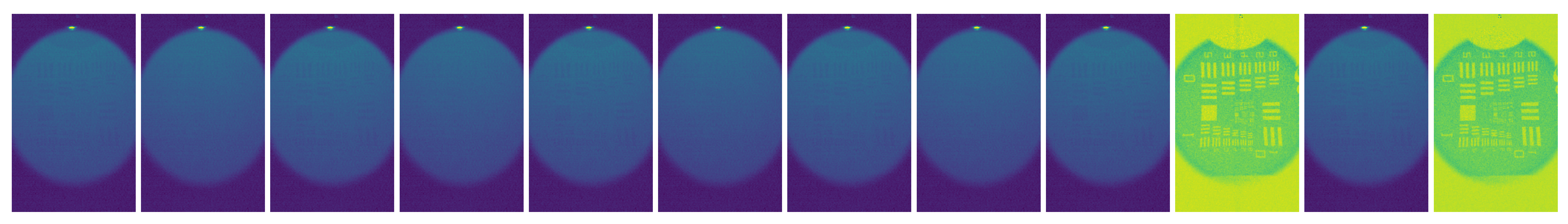}
    \end{subfigure}
    \newline
    \begin{subfigure}[b]{\textwidth}
        \centering
        \includegraphics[scale=0.23]{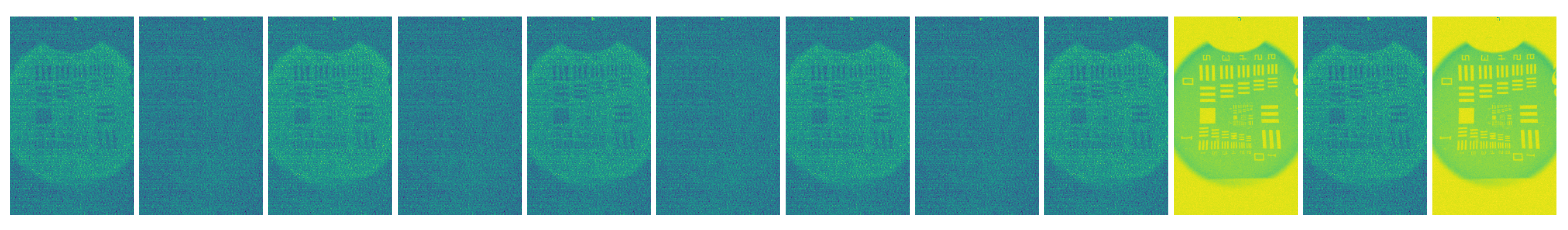}
    \end{subfigure}
    \caption{FPGA-denoised output frames under ambient and noise-free conditions when $G=2$ and $N=4$. Rows show alternating experimental configurations: \textbf{Top row:} Fixed LED adds background noise; \textbf{Bottom row:} No LED interference. In each row, the first $8$ frames represent two PRISM groups ($4$ frames each), while the final $4$ show averaged outputs interleaved with discarded frames. Within the last group, frames $2$ and $4$ illustrate the FPGA denoising kernel’s effectiveness in suppressing LED-induced artifacts and enhancing signal contrast.}
    \label{fig:output-images}
\end{figure}

Figure~\ref{fig:output-images} illustrates the denoising effectiveness of the FPGA preprocessing algorithm. The first row presents output frames in the presence of a background LED acting as a noise source, while the second row depicts results without this interference. In both scenarios, the denoised averages---computed directly on FPGA---successfully suppress transient and ambient noise, yielding enhanced signal clarity. These results were acquired using controlled settings: LED frequency of $200$~Hz, trigger frequency of $400$~Hz, phase offset of $0$, and amplitude of $300$ mV.


\section{Conclusion and Broader Implications}\label{sec:conclusion}
This paper presents a new FPGA-based algorithm that accelerates frame subtraction and averaging, enabling real-time denoising for data generated in PRISM-scale imaging workflows. By leveraging burst-mode optimizations for both DRAM reads and writes, the proposed design outperforms prior approaches that rely on single-mode memory access. Comparative analysis across three algorithmic implementations confirms that this new architecture delivers the highest efficiency and lowest latency. The FPGA kernel is also shown to be scalable---supporting multi-bank processing with multiple FPGAs---without performance degradation. Overall, this implementation dramatically reduces total processing time compared to CPU and GPU workflows, providing a robust solution for latency-sensitive, high-throughput experimental pipelines.


\section*{Acknowledgements}
This research used computational resources provided by the High Performance Computing Center at Michigan State University. The author gratefully acknowledges access to imaging equipment provided by Professor Elad Harel's laboratory (Department of Chemistry, Michigan State University), and acknowledges Professor H. Metin Aktulga (Department of Computer Science and Engineering) for academic guidance during the early stages of this work.

\bibliographystyle{elsarticle-num}
\bibliography{references}

\end{document}